\documentstyle[aps,prbbib,twocolumn,epsf]{revtex}

\begin{document}
\draft
\title{Scattering Matrix Theory For Nonlinear Transport}

\author{Zhong-shui Ma$^{1,2}$, Jian Wang $^1$ and Hong Guo$^{1,3}$}
\address{1. Department of Physics, The University of Hong Kong, 
Pokfulam Road, Hong Kong, China\\
2. Advanced Research Center, Zhongshan University, Guangzhou, China\\
3. Center for the Physics of Materials and Department 
of Physics, McGill University, Montreal, PQ, Canada H3A 2T8\\}
\maketitle

\begin{abstract}
{\bf 
We report a scattering matrix theory for dynamic and nonlinear 
transport in coherent mesoscopic conductors. In general this theory allows
predictions of low frequency linear dynamic conductance, as well as 
weakly nonlinear DC conductance. It satisfies the conditions of gauge 
invariance and electric current conservation, and can be put into a form 
suitable for numerical computation. Using this theory we examine the 
third order weakly nonlinear DC conductance of a tunneling diode.
}
\end{abstract}

\pacs{73.23.Ad,73.40.Gk,72.10.Bg}

\section{Introduction}

Quantum transport under a time-dependent field in coherent 
mesoscopic systems is the subject of many recent 
studies\cite{pas,but1,vegvar,pieper,bruder,chen,wang1}. 
Another problem of interest is the nonlinear conductance 
of such a system, whether under a time-dependent field or
not\cite{but4}.
A difficult theoretical issue is the prediction, for a general 
mesoscopic conductor, the transport coefficients as a function of 
the AC field frequency and the bias voltage. Once these parameters
are known, one can predict useful information such as the nonlinear
current-voltage characteristics in the DC case, the emittance in
the linear frequency, linear voltage AC case, and further nonlinear
dynamic conductance. Indeed, it is now possible to experimentally 
measure the nonlinear AC transport properties such as the second 
harmonic generation, as have been demonstrated by several 
laboratories\cite{webb,vegvar1,liu}.

When a conductor is subjected to time-varying external fields such as 
an AC bias voltage, the total electric current flowing through the
conductor consists of the usual particle current plus the displacement
current. The presence of the latter is crucial such that the total
electric current is conserved. Hence for a theory to deal with AC
transport, in principle one should include the displacement current 
into the consideration. Because a displacement current originates
from induction, and the necessary condition for electric induction is
the electron-electron (e-e) interaction, one thus concludes that an
important ingredient for AC transport theory should be the 
consideration of e-e interactions. These issues have been emphasized
by B\"uttiker and co-workers\cite{review}. 
On the other hand, for DC transport under {\it nonlinear} conditions, 
a necessary requirement is the gauge invariance\cite{but4}: the 
physics should not change when electrostatic potential
everywhere is changed by the same constant amount. Gauge invariance
puts severe conditions on the form of the nonlinear transport
coefficients. From these physical arguments, it is clear that AC as
well DC nonlinear transport contains ingredients which were not
needed when dealing with the familiar DC linear 
transport\cite{review}.

The problems of current conservation and gauge invariance have been 
recognized in the literature. For conductors which maintain quantum
coherence, B\"uttiker and his co-workers have developed an 
approach\cite{but1} based on the single electron scattering matrix theory 
to deal with the linear AC dynamic conductance as well as the 
second order nonlinear conductance coefficients. The original scattering
matrix theory were invented to investigate DC linear transport
coefficients, as is represented by the Landauer-B\"uttiker
formulation\cite{landauer}. Such a theory calculates particle current
from the scattering matrix, thus a direct application to AC situation
would violate current conservation\cite{but1,review}. 
To solve this problem, the scattering matrix theory for AC 
transport consists of two steps\cite{but1,review}. First, it 
calculates the particle current and finds
that this current is not conserved. Second, it considers the e-e
interaction which alters the scattering potential landscape, and this
effect generates an internal response which cancels exactly the
non-conserved part of the particle current thereby restores the current
conservation. For the DC second order nonlinear conductance
coefficient, similar considerations led to the desired gauge
invariance.

In a recent work, the authors have developed a microscopic and general
theoretical formalism for electric response which is appropriate for
both DC and AC weakly nonlinear quantum transport\cite{ma1}.  That
formalism was based on the response theory and it formalized the
connection of the response theory to the scattering matrix theory at
weakly nonlinear level. One of the useful conceptual advances
of the general formalism\cite{ma1} was the introduction of a
frequency dependent characteristic potential at the nonlinear level.  
The characteristic potential describes the changes of scattering 
potential landscape of a mesoscopic conductor when the electrochemical
potential of an electron reservoir is perturbed 
externally\cite{but1}. It is the
nonlinear order characteristic potential which allowed us to analyze
weakly nonlinear AC response\cite{ma1}, as well as the nonlinear DC
conductance, order by order in the bias voltage. In contrast, so far
the scattering matrix theory can be applied up to the second order
nonlinearity and linear order AC.

Using the concept of nonlinear characteristic potential developed in 
the response theory\cite{ma1}, we have found that the scattering 
matrix theory can actually be further developed to apply to 
higher order nonlinear DC situations.  In addition, recognizing that 
an AC transport problem requires the self-consistent solution of the 
Schr\"odinger equation coupled with Maxwell equations, we have found
a way to derive both the external and internal responses in 
{\it equal-footing}
within the scattering matrix approach.  It is the purpose of this
article to report these results. In particular, we shall start from the 
scattering matrix theory and formulate an approach which is appropriate for 
analyzing linear order dynamic conductance and the weakly nonlinear DC 
conductance beyond the second order. We emphasis the properties of electric 
current conservation and gauge invariance, and these properties are maintained
by considering electron-electron interactions. The approach developed
here is particularly useful for nonlinear DC conductance calculation, and we
shall analyze the third order weakly nonlinear transport coefficient for
a double-barrier tunneling diode. Since the approach presented here 
can be cast into a form which allows numerical computation, many further 
applications of it to complicated device structures can be envisioned.

The rest of the article is organized as follows.  In the next section
we present the development of the formalism.  Section III presents two
applications of this formalism: the linear AC dynamic conductance and
the third order nonlinear conductance.  Finally a short conclusion is
included in section IV.

\section{Theoretical Formalism}

In this section we briefly go through the formal development of 
our scattering matrix theory and concentrate more on the conceptually 
important physical quantities which will be needed.

We start by writing the Hamiltonian of the system in the presence of an 
external time-dependent field as
\begin{equation}
H = \sum_{\alpha m} (\bar{E}_{\alpha m} + eV_{\alpha} \cos{\omega t})
a^{\dagger}_{\alpha m}(\bar{E}_{\alpha m},t) a_{\alpha m}(\bar{E}_{\alpha m},t) 
\label{eq1}
\end{equation}
where $a^{\dagger}_{\alpha m}$ is creation operator for a carrier
in the incoming channel $m$ in probe $\alpha$, $eV_{\alpha} \cos{\omega t}$
is the shift of the electrochemical potential $\mu_{\alpha}$ away from
the equilibrium state associated with $\mu^{eq}$, {\it i.e.}, $eV_{\alpha} 
\cos{\omega t} = \mu_{\alpha} - \mu^{eq}$. The energy $\bar{E}_{\alpha m}$
is a functional of the internal electrical potential landscape
$U({\bf r},\{V_{\alpha}\})$ which depends on $V_{\alpha}$ in the low 
frequency regime. Potential $U$ includes the internal response to
the external perturbation and it generates such effect as the displacement
current. In general $U$ is also an explicit function of time (or of the
AC frequency $\omega$) as discussed in Ref. \onlinecite{ma1}, but in this
work we shall only be concerned with the dynamic conductance to first
power of $\omega$ and for this case $U$ is static. Note we have
explicitly included $U$ into the Hamiltonian which helps in dealing with
both external and internal responses in equal-footing.
The self-consistent nature of this Hamiltonian is clear: $U$ must be
determined, in general, from the Maxwell equations where the charge
density is obtained from solving the quantum mechanical problem of the
Hamiltonian.

Next let's consider the series expansion of the energy in terms of the 
potential landscape $U$,
\[\bar{E}_{\alpha m}+eV_{\alpha}cos(\omega t) = 
E_{\alpha m} + e\hat{O}_{\alpha}^{(1)}
cos(\omega t)\] 
\begin{equation}
+ e^2\hat{O}_{\alpha}^{(2)}(cos (\omega t))^2 +
\cdot\cdot\cdot
\label{eq2}
\end{equation}
where the operators $\hat{O}_{\alpha}^{(i)}$ is a spatial integration of
the $i$-th order {\it characteristic potential} (see below) folded with 
the $i$-th order functional derivative of $E_{\alpha m}$ with respect to 
the potential landscape $U({\bf r})$. For instance the linear order operator,
which is linear in voltage $V_{\beta}$, is given by
\begin{equation}
\hat{O}_{\alpha}^{(1)} \equiv \sum_{\beta} (\delta_{\alpha \beta}
+\partial_{V_{\beta}} E_{\alpha m}) V_{\beta}
\label{eq5}
\end{equation}
where 
\[\partial_{V_{\beta}}E\ \equiv \ \int d^3 {\bf r}\ u_{\beta}({\bf r})
\ \frac{\delta E}{\delta eU({\bf r})} \]
with $u_{\beta}({\bf r})\equiv \frac{\partial U({\bf r})}{\partial V_{\beta}}$
the linear order characteristic potential\cite{but1}. The expressions
for higher order operators $\hat{O}_{\alpha}^{(i)}$ are more
difficult to write down in a general form, but they are proportional
to the $i$-th power of the bias. In addition they can be easily 
determined after we formally obtain the transmission function and 
then applying the current conservation and gauge invariance to the 
results. Using Eq.(\ref{eq2}), the Hamiltonian now reads
\begin{eqnarray}
H &=& \sum_{\alpha m} \left[E_{\alpha m} + \sum_i\hat{O}_{\alpha}^{(i)}
(cos{\omega t})^i\right] \nonumber \\
& &\times
a^{\dagger}_{\alpha m}(\bar{E}_{\alpha m},t) 
a_{\alpha m}(\bar{E}_{\alpha m},t)\ \ \ .
\label{eq4}
\end{eqnarray}
The operators $a_{\alpha m}(\bar{E}_{\alpha m},t)$ satisfies the equation 
of motion 
\begin{equation}
\dot{a}_{\alpha m}(\bar{E}_{\alpha m},t) = \frac{1}{i\hbar} [a_{\alpha m}
(\bar{E}_{\alpha m},t), H]\ \ ,
\label{eq6}
\end{equation}
which can be integrated because the time dependence of $H$ is simple. 
For instance to linear order in voltage, we only need to use
$\hat{O}^{(1)}$ in the Hamiltonian and the result is
\begin{eqnarray}
a_{\alpha,m}(\bar{E}_{\alpha m},t)&=& 
a_{\alpha m}(\bar{E}_{\alpha m}) \nonumber \\
&\times & \exp\left(-\frac{i}{\hbar} \left[E_{\alpha m} t +
\frac{e\hat{O}_{\alpha}^{(1)}}{\omega} \sin{\omega t}\right]\right)
\nonumber
\end{eqnarray}
Its Fourier transform is given by
\begin{eqnarray}
\tilde{{\bf a}}_{\alpha}(E) & & =
\int dt {\bf a}_{\alpha}(\bar{E}_{\alpha},t) e^{i Et/\hbar}
= {\bf a}_{\alpha}(E) \nonumber\\
& & -\frac{e}{2\hbar \omega} \hat{O}_{\alpha}^{(1)}
[{\bf a}_{\alpha}(E+\hbar \omega) - 
{\bf a}_{\alpha}(E-\hbar \omega)]\nonumber\\
& & +\frac{e^2}{8\hbar^2 \omega^2} 
\left(\hat{O}_{\alpha}^{(1)}\right)^2 \left[{\bf a}_{\alpha}
(E+2\hbar \omega) -2 {\bf a}_{\alpha}(E)\right.\nonumber\\
& &\left. + {\bf a}_{\alpha}(E-2\hbar \omega)\right]
+ \cdot\cdot\cdot\nonumber\\
& & = \sum_n \frac{1}{n!} (
\frac{-e\hat{O}_{\alpha}^{(1)}}{2\hbar \omega})^n
(e^{\hbar \omega \partial_E} - e^{-\hbar \omega \partial_E})^n
{\bf a}_{\alpha}(E)
\label{eq8}
\end{eqnarray}
where we have suppressed the index $m$ and ${\bf a}_{\alpha}$ 
is in a vector form of the operators $a_{\alpha m}$. 
In Eq.(\ref{eq8}) the physics is transparent: 
${\bf a}_{\alpha}(E \pm \hbar \omega)$ is just the one-photon sideband and 
${\bf a}_{\alpha}(E \pm 2\hbar \omega)$ corresponds to the second
harmonic generation. More tedious expressions can be obtained
if higher order operators $\hat{O}^{(i)}$ are included in the
Hamiltonian.

To calculate the total electrical current, we shall apply the formula 
derived in Ref. \onlinecite{but2}, which is exact up to linear order of 
$\omega$ and for larger frequency it is an approximation to a 
space-dependent expression of the current operator,
\begin{eqnarray}
I_{\alpha}(t)&=& \frac{e}{h} \int dE dE' \left[ \tilde{{\bf
a}}_{\alpha}^{\dagger}(E) \tilde{{\bf a}}_{\alpha}(E')
\right.\nonumber\\
& -&\left. \tilde{{\bf b}}_{\alpha}^{\dagger}(E) 
\tilde{{\bf b}}_{\alpha}(E') \right]\times \exp(i(E-E')t/\hbar)
\label{eq9}
\end{eqnarray}
where $\tilde{{\bf b}}_{\alpha}(E)$ is the operator which annihilates a 
carrier in the outgoing channel in probe $\alpha$. The annihilation 
operator in the outgoing channel, $\tilde{{\bf b}}_{\alpha}$, is related 
to the annihilation operator in the 
incoming channel $\tilde{{\bf a}}_{\alpha}$ via the scattering matrix
${\bf s}_{\alpha\beta}$: $\tilde{{\bf b}}_{\alpha} 
= \sum_{\beta} {\bf s}_{\alpha \beta} \tilde{{\bf a}}_{\beta}$
where ${\bf s}_{\alpha\beta}$ is a function of energy $E$ and a functional
of the electric potential $U({\bf r},\{V_{\alpha}\})$.
Finally we comment that in evaluating Eq. (\ref{eq9}) we need to take
a quantum statistical average of 
$<{\bf a}_{\alpha}^{\dagger}(E) {\bf a}_{\beta}(E')> = \delta_{\alpha
\beta} \delta(E-E') f_{\alpha}(E)$
where $f_{\alpha}(E)$ is the Fermi function of reservoir $\alpha$.
Because of the limitations of Eq. (\ref{eq9}), our theory will 
be exact for transport coefficients linear in $\omega$ for AC situations.
However this is not a severe limitation for practical 
calculations\cite{but2}.

One of the most important quantities of this theory is the determination
of characteristic potential which arrives naturally. As discussed 
above, this quantity determines the operators $\hat{O}^{(i)}$. 
Since the scattering matrix theory used here is exact to linear power of
$\omega$ which is the order we shall work on, we only need to consider
$\omega$-independent characteristic potentials.  On the other hand,
as we are interested in the weakly nonlinear coefficients, it is
crucial to consider higher order characteristic potentials\cite{ma1}: 
$u_{\delta\gamma}({\bf r})\equiv \partial^2 U({\bf r})/\partial
V_{\delta}\partial V_{\gamma}$, $u_{\beta\delta\gamma}({\bf r})$, etc.. 
For any physical quantity beyond the terms linear in $\omega$ or second 
order in voltage, including the second harmonic generation term (the term 
of $\omega V^2$), these higher order characteristic potentials are
necessary. 

We now discuss the solution of higher order characteristic potentials
by explicitly carrying out the calculation of $u_{\delta\gamma}$. In the
weakly nonlinear regime, the variation of the electric potential can be 
expanded in terms of the variation of the electrochemical potential $d\mu$ 
\begin{equation}
edU({\bf r}) = \sum_{\beta} u_{\beta}({\bf r}) d\mu_{\beta} 
+ \frac{1}{2} \sum_{\beta \gamma} u_{\beta \gamma}({\bf r}) 
d\mu_{\beta} d\mu_{\gamma} + ...
\label{dU}
\end{equation}
where $u_{\beta}$ is the characteristic potential, $u_{\beta \gamma}$
(which is symmetric in $\beta$ and $\gamma$) is the second order
characteristic potential tensor, and $(\cdot\cdot\cdot)$ are higher
order terms written in a similar fashion. Because we are only 
interested in AC transport to the first power of frequency
$\omega$, the electrodynamics is solved by the Poisson equation,
\begin{equation}
-\nabla^2 dU({\bf r}) = 4\pi e^2dn({\bf r}) 
= 4\pi e^2\sum_{\alpha} dn_{\alpha}({\bf r})
\label{poisson}
\end{equation}
where $dn_{\alpha}$ is the variation of the charge density at contact 
$\alpha$ due to a change in electrochemical potential at that contact. 
There are two contributions to the charge density at contact $\alpha$: 
the injected charge density due to the variation of the chemical 
potential at contact $\alpha$, and the induced charge density
$dn_{ind,\alpha}$ due to the electro-static potential, hence
\begin{equation}
dn_{\alpha} = \frac{dn_{\alpha}}{dE} d\mu_{\alpha} + \frac{1}{2}
\frac{d^2n_{\alpha}}{dE^2} d\mu_{\alpha}^2 +\cdot\cdot\cdot\ +\  
dn_{ind,\alpha}
\label{induce}
\end{equation}
where $dn_{\alpha}/dE$ is the injectivity which is the local density of
state at contact $\alpha$ and $d^2n_{\alpha}/dE^2$ is the energy
derivative of the injectivity. The induced charge density involves Lindhard 
function. Using Thomas-Fermi approximation Eq.(\ref{induce}) takes a 
compact and simple form
\begin{equation}
dn_{\alpha} = \frac{dn_{\alpha}}{dE} (d\mu_{\alpha} - edU) + \frac{1}{2}
\frac{d^2n_{\alpha}}{dE^2} (d\mu_{\alpha}-edU)^2\ +\ \cdot\cdot\cdot
\label{dn}
\end{equation}
From Eqs.(\ref{dU}), (\ref{poisson}), and ({\ref{dn}), we obtain the
equation satisfied by the second order characteristic potential tensor
\begin{eqnarray}
-\nabla^2 u_{\beta \gamma} + 4\pi e^2 \frac{dn}{dE} u_{\beta \gamma}
&=& 4\pi e^2 \left(\frac{d^2n_{\beta}}{dE^2} \delta_{\beta \gamma} -
\frac{d^2n_{\gamma}}{dE^2} u_{\beta}\right. \nonumber \\
&-& \left.\frac{d^2n_{\beta}}{dE^2} u_{\gamma} + \frac{d^2n}{dE^2} u_{\beta}
u_{\gamma}\right)\ .
\label{u11}
\end{eqnarray}
Since all the quantities involved in this equation are known from the 
linear order calculation, $u_{\beta\gamma}$ can thus be determined. 
Similarly,
order by order we can determine higher order characteristic potentials
from results obtained at lower orders, for instance the equation
satisfied by $u_{\beta\gamma\delta}$ is found to be:
\begin{eqnarray}
& & -\nabla^2 u_{\beta \gamma \delta} + 4\pi e^2 
\frac{dn}{dE} u_{\beta \gamma \delta} = 4\pi e^2 
\left(\frac{d^3n_{\beta}}{dE^3} \delta_{\beta \gamma} 
\delta_{\beta \delta} \right.\nonumber\\
& & - \frac{d^3n}{dE^3} u_{\beta} u_{\gamma} u_{\delta}
+ \left\{ \frac{d^3n_{\beta}}{dE^3} u_{\gamma} u_{\delta}
- \frac{d^3n_{\beta}}{dE^3} \delta_{\beta \gamma} u_{\delta}
\right.\nonumber\\
& &\left.\left. + \frac{d^3n}{dE^3} u_{\beta \gamma} u_{\delta} 
- \frac{d^3n_{\beta}}{dE^3} u_{\gamma \delta} \right\}_c \right)
\label{u111}
\end{eqnarray}
where the curl bracket $\{ ... \}_c$ stands for the cyclic permutation of 
indices $\beta$, $\gamma$, and $\delta$.
Note that if better models are needed to deal with the
screening effect, the term with $dn/dE$ on the left hand side of 
(\ref{u11}), (\ref{u111}) and higher order equations 
is replaced by an integration over the appropriate
Lindhard function folded with the characteristic potential\cite{ma1}.

From Eq.(\ref{dU}) we can derive several important sum rules on the 
characteristic potential tensor. If all the changes in the 
electrochemical potentials are the same, {\it i.e.},
$d\mu_{\beta} = d\mu_{\gamma} = d\mu$, this corresponds to an overall 
shift of the electro-static potential $edU = d\mu$. From this we have 
$\sum_{\beta} u_{\beta} = 1$, $\sum_{\beta \gamma} u_{\beta \gamma} = 0$.
Due to gauge invariance, Eq.(\ref{dU}) remains the same if $dU$, 
$d\mu_{\beta}$, and $d\mu_{\gamma}$ are all shifted by the same amount. 
This leads to 
$\sum_{\beta} u_{\beta \gamma} = \sum_{\gamma} u_{\beta \gamma} = 0$.
Using Eq.(\ref{u11}) we can confirm that these relations 
are indeed satisfied. Similar sum rules can be derived for higher order
characteristic potentials.

Let's summarize the scattering matrix theoretical procedure. With the
characteristic potential tensor calculated, we explicitly derive 
the Hamiltonian in a series form Eq. (\ref{eq4}). The Hamiltonian 
determines the creation and annihilation operators via equation of
motion Eq. (\ref{eq6}) and the scattering matrix 
${\bf s}_{\alpha\beta}$. Finally, using Eq. (\ref{eq9}) we compute the
electric current as a function of voltage. 

\section{Applications}

In the following we apply the scattering matrix formalism developed in
the last section to two examples: the linear order emittance and the third
order DC nonlinear conductance. The first example has been examined by 
B\"uttiker and co-workers\cite{but1}, our result is in exact agreement 
with theirs. The second example has not been studied and we shall
provide further numerical results for a resonant tunneling diode.

\subsection{Linear Dynamic Conductance}

The linear dynamic conductance (called emittance) is the transmission 
function of the terms proportional to $V_\beta$ and $\omega V_{\beta}$ 
in the electric current. From Eq. (\ref{eq9}) we expand every thing in 
these variables and obtain

\begin{eqnarray}
I_{\alpha}(\omega') &=& \sum_{\beta} \hat{O}_{\beta}^{(1)} \int dE 
(-\partial_E f) \{ \frac{e^2}{h} A_{\beta \beta}(\alpha,E,E) \nonumber \\
& -& \frac{e^2}{2h} \hbar \omega' [{\bf s}_{\alpha \beta}^{\dagger} 
\partial_E {\bf s}_{\alpha \beta} - (\partial_E {\bf s}_{\alpha \beta}) 
{\bf s}_{\alpha \beta}] \}
\label{eq16}
\end{eqnarray}
where we have used notation $A_{\alpha \beta} \equiv A_{\beta
\beta}(\alpha,E,E)$ and ${\bf s}_{\alpha \beta} \equiv {\bf s}_{\alpha
\beta}(E,U)$. The transmission function $A$ is defined in the usual 
form as
\begin{eqnarray}
A_{\beta \beta'}(\alpha, E, E',U)& =& 
{\bf 1}_{\alpha} \delta_{\alpha \beta} \delta_{\alpha \beta'}
\nonumber\\
& -& {\bf s}_{\alpha \beta}^{\dagger}(E,U)
{\bf s}_{\alpha \beta'}(E',U)
\label{eq12}\ \ .
\end{eqnarray}
In deriving Eq. (\ref{eq16}), we have used the fact that 
$\sum_{\beta} A_{\beta \beta} (\alpha, E,E,U)=0$. All quantities 
such as $A_{\alpha \beta}$, ${\bf s}_{\alpha \beta}$, and 
$u_{\beta}=(\partial U({\bf r})/\partial V_{\beta})_{eq}$ 
are taken at equilibrium, {\it i.e.}, at $V_{\alpha}=0$. 
Using Eq.(\ref{eq5}), we separate the operator according to 
$\hat{O}_{\beta}^{(1)}
\partial_E = V_{\beta} \partial_E + \sum_{\gamma} V_{\gamma} 
\partial_{V_{\gamma}}$, thus (\ref{eq16}) can further be simplified to,
\begin{eqnarray}
I_{\alpha}(\omega') &=& \int dE (-\partial_E f) \{ \sum_{\beta} V_{\beta} 
\frac{e}{h} A_{\alpha \beta} \nonumber \\
& -& \frac{e^2}{2h} \hbar \omega' \sum_{\beta} V_{\beta}
[{\bf s}_{\alpha \beta}^{\dagger} \partial_E {\bf
s}_{\alpha \beta} - (\partial_E {\bf s}_{\alpha \beta}) {\bf
s}_{\alpha \beta}] \nonumber \\
& -& \frac{e^2}{2h} \hbar \omega' \sum_{\gamma} V_{\gamma} 
\sum_{\beta} [{\bf s}_{\alpha \beta}^{\dagger}
\partial_{V_{\gamma}} {\bf s}_{\alpha \beta} - (\partial_{V_{\gamma}} 
{\bf s}_{\alpha \beta}) {\bf s}_{\alpha \beta}] \}
\label{eq17}
\end{eqnarray}
From this result, we immediately realize that the first term on the right 
hand side is just the DC contribution to the electric current. From the 
second and third terms which are linear in $\omega'$ and $V_{\beta}$, 
we obtain the linear order emittance
\begin{eqnarray}
E_{\alpha \beta} &=& \int dE (-\partial_E f) \frac{e^2}{4\pi i} \{ 
[{\bf s}_{\alpha \beta}^{\dagger} \partial_E {\bf s}_{\alpha \beta} 
- (\partial_E {\bf s}_{\alpha \beta}) {\bf s}_{\alpha \beta}] 
\nonumber \\
&+& \sum_{\gamma}[{\bf s}_{\alpha \gamma}^{\dagger} \partial_{V_{\beta}} 
{\bf s}_{\alpha \gamma} - (\partial_{V_{\beta}} 
{\bf s}_{\alpha \gamma}) {\bf s}_{\alpha \gamma}] \}\ \ .
\label{eq18}
\end{eqnarray}
This result exactly agrees with the that obtained previously\cite{but1}. 
The first term of $E_{\alpha\beta}$ describes the external contribution
to the AC current, while the second term is from internal response.
They are obtained simultaneously from the scattering matrix theory
developed here. Finally, from the gauge invariance condition\cite{but1}
$e\partial_E A_{\alpha \beta} + \sum_{\gamma} \partial_{V_{\gamma}}
A_{\alpha \beta} = 0$, it is easy to show 
$\sum_{\beta} E_{\alpha \beta} = 0$, which is a direct consequence of 
$\hat{O}_{\beta}^{(1)}$ in Eq. (\ref{eq16}). It is
also easy to show that the electric current is conserved, {\it i.e.}, 
$\sum_{\alpha} E_{\alpha \beta} = 0$.

\subsection{Third Order DC Nonlinear Conductance}

The scattering matrix theory developed here can be applied to compute 
DC weakly nonlinear conductance to any order in bias. As an example we 
now calculate the third order DC nonlinear conductance
$G_{\alpha \beta \gamma \delta}$, which is defined by expanding the
electric current in powers of voltage to the third power,
\begin{eqnarray}
I_{\alpha}& =& \sum_{\beta} G_{\alpha \beta}V_{\beta} + 
\sum_{\beta \gamma} G_{\alpha \beta \gamma}V_{\beta}V_{\gamma}
\nonumber\\
&+& \sum_{\beta \gamma \delta} G_{\alpha \beta \gamma \delta}
V_{\beta}V_{\gamma}V_{\delta} +\ \cdot\cdot\cdot\ \  .
\end{eqnarray}
Following the same procedure as above in deriving the linear
emittance $E_{\alpha \beta}$, by expanding the electric current Eq.
(\ref{eq9}) and other quantities to third order in bias,
it is tedious but straightforward to derive 
\begin{eqnarray}
G_{\alpha \beta \gamma \delta} &=& \frac{e^3}{3h} 
\int dE (-\partial_E f) \left[\left\{\partial_{V_{\gamma}}
\partial_{V_{\delta}} A_{\alpha \beta}\right.\right.\nonumber\\
&+&\left.\left. e\partial_{V_{\gamma}} \partial_E A_{\alpha \beta} 
\delta_{\beta \delta}
\right\}_c + e^2\partial_E^2 A_{\alpha \beta} 
\delta_{\beta \gamma} \delta_{\gamma \delta}\right]\ .
\label{g4}
\end{eqnarray}
Note that the second order characteristic potential 
tensor has been implicitly included in Eq.(\ref{g4}), because
\begin{eqnarray}
\partial_{V_{\gamma}} \partial_{V_{\delta}} A_{\alpha \beta} &=& 
\partial_{V_{\gamma}} \int \frac{\delta A_{\alpha \beta}}{\delta
U({\bf r}_1)} \frac{\partial U({\bf r}_1)}{\partial V_{\delta}}
d{\bf r}_1 \nonumber \\
&=& \int \left(\frac{\delta}{\delta U({\bf r}_2)}
\frac{\delta A_{\alpha \beta}}{\delta U({\bf r}_1)}\right)
\ u_{\delta}\ u_{\gamma}\ d{\bf r}_1 d{\bf r}_2 \nonumber \\
&+& \int \frac{\delta A_{\alpha \beta}}{\delta U({\bf r}_1)} 
\ u_{\delta \gamma}\ d{\bf r}_1 \ \ .
\label{relation}
\end{eqnarray}
Again, we emphasis that other higher order nonlinear conductance 
can be calculated in a similar fashion. In general, the n-th order 
characteristic potential tensor is needed for the $(n+1)$th order 
nonlinear conductance.  Finally we point out that this result
Eq.(\ref{g4}) can in fact be obtained by expanding the following electric 
current expression to the third order in voltage,
$I_{\alpha} = \frac{2e}{h} \sum_{\beta} \int dE f(E-E_F-eV_{\beta})
A_{\alpha \beta}(E,\{V_{\gamma}\})$.

In the following we calculate $G_{1111}$ from the general result of Eq.
(\ref{g4}) for a double barrier tunneling diode. For simplicity let's 
consider an one-dimensional double barrier tunneling system 
where the two barriers are $\delta$-functions located 
at positions $x=-a$ and $x=a$. The barrier strength is $V_1$ and $V_2$
respectively. When $V_1=V_2$, this is a symmetric system, hence the 
second order nonlinear conductance vanishes. In the symmetric case the 
first nonlinear coefficient comes from the third order, as specified by 
Eq. (\ref{g4}). If we approximate the scattering matrix by the Breit-Wigner 
form\cite{but4} near a resonance energy $E_r$,
$s_{\alpha\beta}(E)\sim
[\delta_{\alpha\beta}-i\sqrt{\Gamma_\alpha\Gamma_\beta}/\Delta]$,
where $\Gamma_\alpha$ is the decay width of barrier $\alpha$,
$\Delta=\Delta E+i\Gamma/2$ with $\Gamma=\Gamma_1+\Gamma_2$ and
$\Delta E\equiv E-E_r$, we obtain a simple expression,
\begin{eqnarray}
& & G_{1111} = \frac{2e^3}{3h{\Gamma}^2}[(3(\Delta E)^2-
\frac{\Gamma^2}{4})(\Gamma_1^2+\Gamma_2^2-\Gamma_1 \Gamma_2) 
\Gamma_1 \Gamma_2\nonumber\\
& &-6\Gamma_1^2 \Gamma_2^2 (\Delta E)^2]
\left[(\Delta E)^2+\frac{\Gamma^2}{4}\right]^{-3}
\label{breit}
\end{eqnarray}
For the symmetric case, this expression reduces to
\begin{equation}
G_{1111} = -\frac{e^3 \Gamma_1^2}{6 h} \frac{3(\Delta E)^2+
\Gamma_1^2}{[(\Delta E)^2 + \Gamma_1^2]^3}
\label{breit1}
\end{equation}
which is negative definite and has one minimum at $\Delta E = 0$. 
Because of the simple nature of the scattering matrix within the
Breit-Wigner form, a general electric current expression has been
obtained\cite{but4}: $I_{\alpha}=I_{\alpha}(V_{\beta})$. We have 
thus calculated $G_{1111}$ from this exact I-V relation, and it 
agrees exactly with the result (\ref{breit1}) which comes from 
Eq. (\ref{g4}).

Most practical transport problems can not be solved analytically. It is 
thus very important to be able to solve them numerically.
Indeed, a distinct merit of the scattering matrix theory presented here
is that it allows numerical computation, {\it e.g.} Eqs. (\ref{eq18}), 
and (\ref{g4}) can be numerically evaluated for explicit 
scattering potentials of a conductor. We only mention that the functional 
derivatives of the transmission function $A_{\alpha\beta}$ with respect 
to the potential landscape $U({\bf r})$ as appeared in Eq. (\ref{g4}), the 
potential derivatives, and the partial local density of states which is 
needed in the emittance calculation, can all be determined via
the scattering Green's function. 

For the double-barrier diode just discussed, if we do not use the 
Breit-Wigner scattering form, $G_{1111}$ can only be obtained numerically.  
For this system the Green's function $G(x,x')$ can be calculated 
exactly\cite{yip} thus from the Fisher-Lee relation\cite{lee} 
we obtain the scattering matrix ${\bf s}_{\alpha\beta}(E)$ and hence the 
transmission function $A_{\alpha\beta}$ from its definition (\ref{eq12}).
To compute $\delta A_{11}/\delta U$ and the higher order functional
derivative, we use Fisher-Lee relation and the fact\cite{gasparian}
that $\delta G(x_1,x_2)/\delta U(x) = G(x_1,x) G(x,x_2)$.
Hence,
\begin{equation}
\frac{\delta {\bf s}_{11}}{\delta U(x)} = i\hbar v G(x_1,x) G(x,x_1)
\end{equation}
and
\begin{eqnarray}
& & \frac{\delta^2 {\bf s}_{11}}{\delta U(x) \delta U(x')} 
=i\hbar v \left[G(x_1,x') G(x',x) G(x,x_1)\right.\nonumber\\
& & \left. + G(x_1,x) G(x,x') G(x',x_1)\right]\ \ ,
\end{eqnarray}
where $v$ is the velocity of the particle. The energy and
$V_\gamma$ derivatives of Eq. (\ref{g4}) can be evaluated explicitly
using the numerical procedures documented before\cite{wang1,wang2}. 
Finally, the nonlinear characteristic potential $u_{11}$ is obtained 
from Eq. (\ref{u11}).

\begin{figure}
\narrowtext
{\epsfxsize=7cm\epsfysize=7.0cm\centerline{\epsfbox{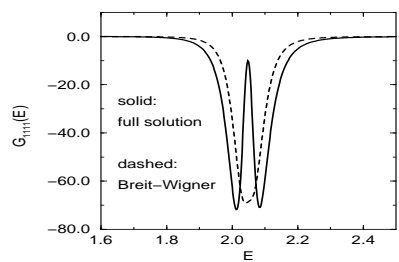}}}
\vspace{-2.5cm}
\caption{
$G_{1111}$ as a function of the scattering electron energy $E$ for a
double barrier tunneling diode with symmetrical barriers.  Solid line:
numerical results by solving the full quantum scattering problem using
Green's functions.  Dashed curve: using the approximate Breit-Wigner
form of the scattering matrix. The units of the quantities are set by
$\hbar=1$, $e=1$ and $m=1/2$.
}
\label{fig1}
\end{figure}

The numerical result for $G_{1111}$ as a function of the scattering electron
energy $E$ is plotted as the solid curve in Fig. (1). Around the resonance
energy $E_r\approx 2.0$, $G_{1111}$ takes large but negative values.
Thus for electron Fermi energy in this range, the current-voltage
characteristics will be nonlinear and this may result to negative
differential resistance. Such a behavior has important significance for
practical purposes\cite{datta}. Notice that the two dips of $G_{1111}$ are
not exactly the same, such an asymmetrical behavior has been observed in 
other quantities\cite{yip}. For comparison we also plotted (dashed line)
the result from the Breit-Wigner approximation of the scattering matrix, Eq.
(\ref{breit1}). While the negative nature of $G_{1111}$ and the overall
magnitude are similar to the numerical result, the Breit-Wigner result
shows only one dip.  This inconsistency is completely due to the simple
form of the Breit-Wigner approximation as it gives a space-independent
and constant characteristic potential. The accurate solution using the
Green's function generates space-dependence of various quantities.

\section{Summary}

In summary, we have extended the scattering matrix theory which is
now appropriate to analyzing linear dynamic conductance to first order in
frequency, and weakly nonlinear DC conductance order by order in
external bias. The crucial ingredient of this development is the
characteristic potential at weakly nonlinear orders and these potentials
appear naturally from the self-consistent Hamiltonian. The theory is 
current conserving and gauge invariant. The physical quantities involved 
in this theory are numerically calculable, hence the present approach
can be used to conductors with complicated scattering potential landscape for 
quantitative predictions. The formal connection of the scattering 
matrix theory and the response theory, at the weakly nonlinear AC level, 
has been clarified in our recent work\cite{ma1}. While the response theory 
is very general\cite{ma1} and can be used to analyze weakly nonlinear DC and 
AC transport order by order in bias, and for AC case to all orders of 
frequency, we expect that the scattering matrix theory should be able to 
do the same. This paper partially fulfills this expectation by
extending the scattering matrix theory to higher orders of nonlinearity.
We point out that to go beyond the linear frequency, the expression for
electric current, Eq. (\ref{eq9}), should be extended. In addition we
should use the Helmhotz equation (in Lorentz gauge) for the 
electrodynamics instead of the Poisson equation. 
Finally, it is important to note that the theoretical 
approach of scattering matrix developed here is appropriate to transport 
problems near equilibrium. Far from equilibrium, one may employ the Keldysh 
Green's functions\cite{keldysh}.

\noindent
{\bf Acknowledgments}. 
We thank Mr. Q.R. Zheng for a useful discussion concerning the 
electrodynamics used in this work. We gratefully acknowledge financial
support by various organizations. J.W. is supported by a research 
grant from the Croucher Foundation, a RGC grant from the SAR 
Government of Hong Kong under grant number HKU 261/95P, and a 
CRCG grant from the University of Hong Kong; Z.M. is supported by 
the Foundation of Advanced Research Center of Zhongshan University 
and NSF-China; H.G. is supported by NSERC of Canada and FCAR 
of the Province of Qu\'ebec. We thank the Computing Center of The 
University of Hong Kong for CPU allocations.

\end{document}